\documentclass[10pt,aps,pre,twocolumn,superscriptaddress,showkeys,preprintnumbers]{revtex4-1}

\usepackage{amsmath,amssymb,amsfonts}
\usepackage{graphicx,color}
\usepackage{dcolumn}
\usepackage{bm}
\usepackage{hyperref}
\usepackage{braket}
\usepackage{dsfont}
\usepackage{float}

\newcommand{\ie}{\emph{i.e.}}
\newcommand{\eg}{\emph{e.g.}}
\newcommand{\ea}{\emph{et al.} }
\newcommand{\ER}{Erd\"os-R\'enyi }
\newcommand{\SF}{scale-free }

\begin{document}

\title{Multiple structural transitions in interacting networks} 

\author{Giacomo Rapisardi}
\email{giacomo.rapisardi@imtlucca.it}
\affiliation{IMT School for Advanced Studies, 55100 Lucca, Italy}
\author{Alex Arenas}
\affiliation{Departament d'Enginyeria Inform\`atica i Matem\`atiques, Universitat Rovira i Virgili, 43007 Tarragona, Spain}
\author{Guido Caldarelli}
\affiliation{IMT School for Advanced Studies, 55100 Lucca, Italy}
\affiliation{Istituto dei Sistemi Complessi (ISC)-CNR, 00185 - Rome, Italy}
\author{Giulio Cimini}
\affiliation{IMT School for Advanced Studies, 55100 Lucca, Italy}
\affiliation{Istituto dei Sistemi Complessi (ISC)-CNR, 00185 - Rome, Italy}
\date{\today}

\begin{abstract}
Many real-world systems can be modeled as interconnected multilayer networks, namely a set of networks interacting with each other. 
Here we present a perturbative approach to study the properties of a general class of interconnected networks as inter-network interactions are established. 
We reveal multiple structural transitions for the algebraic connectivity of such systems, between regimes in which each network layer keeps its independent identity 
or drives diffusive processes over the whole system, thus generalizing previous results reporting a single transition point. 
Furthermore we show that, at first order in perturbation theory, the growth of the algebraic connectivity of each layer depends only 
on the degree configuration of the interaction network (projected on the respective Fiedler vector), and not on the actual interaction topology. 
Our findings can have important implications in the design of robust interconnected networked system, 
particularly in the presence of network layers whose integrity is more crucial for the functioning of the entire system. 
We finally show results of perturbation theory applied to the adjacency matrix of the interconnected network, 
which can be useful to characterize percolation processes on such systems.
\end{abstract}
\maketitle

\section{Introduction}

Interconnected (or interdependent) networks describe complex systems composed by a set of networks interacting with each other 
\cite{dedomenico2013mathe,kivela2014multilayer,boccaletti2014structure,garas2016interconnected}. 
The presence of such interactions makes these systems structurally and dynamically different from isolated networks \cite{radicchi2014driving,dedomenico2016physics}. 
Dissimilar properties have been reported, for instance, in navigability \cite{dedomenico2014navigability}, communicability \cite{estrada2014communicability},
robustness \cite{buldyrev2010catastrophic,gao2011robustness,gomezgardenes2015layer}, percolation \cite{hu2011percolation,bianconi2014multiple,hackett2016bond}, 
epidemics \cite{saumell2012epidemic,dickison2012epidemics,wang2012epidemics,granell2013dynamical}, and synchronization \cite{huang2006abnormal,aguirre2014syncronization}. 

In the study of interconnected networks, much attention has been devoted to the Laplacian operator 
\cite{radicchi2013abrupt,gomez2013diffusion,sole2013spectral,martinez2014algebraic,sanchez2014dimensionality,darabi2015exact,shakeri2016maximizing,van2016interconnectivity}. 
The Laplacian matrix $\mathcal{L}$ of an undirected graph is defined as $\mathcal{D}-\mathcal{A}$, 
where $\mathcal{A}$ is the adjacency matrix (its generic element $A_{ij}=1$ if $i$ and $j$ are connected, and $A_{ij}=0$ otherwise) 
and $\mathcal{D}=\mbox{diag}(\mathcal{A}\ket{1})$ is the diagonal matrix of degrees (we use the bra-ket notation, hence $\ket{1}$ denotes the column vector with all entries equal to 1).  
$\mathcal{L}$ is positive semidefinite, meaning that all of its eigenvalues are non-negative. 
Since, by definition, row/column sums of $\mathcal{L}$ are all zero, the Laplacian always admits $\lambda_1(\mathcal{L})=0$ as the smallest eigenvalue, corresponding to the eigenvector $\ket{1}$.  
The second-smallest eigenvalue of the spectrum, $\lambda_2(\mathcal{L})$, is the \emph{algebraic connectivity} of the graph, and reflects how much connected the overall graph is \cite{fieldedr1975prop}. 
Indeed, $\lambda_2(\mathcal{L})$ is different from zero if and only if the graph is connected; otherwise, its degeneracy equals the number of disconnected components of the graph. 
The value of $\lambda_2(\mathcal{L})$ is determined as:
\begin{equation}\label{eqn:lambda_courant}
\lambda_2 (\mathcal{L}) = \min_{\ket{v} \in \mathsf{V}} \braket{v| \mathcal{L} |v}
\end{equation} 
where $\ket{v} \in \mathsf{V}$ is such that $\braket{v|1}=0$ and $\braket{v|v} = 1$.

The spectrum of the graph Laplacian is typically used to characterize both structural properties of the networked system, 
such as connectivity, diameter and number of spanning trees \cite{mohar1991laplacian,jamakovic2008robustness}, 
as well as dynamical properties, such as diffusion and synchronization \cite{almendral2007dynamical,arenas2008syncronization,samukin2008laplacian}. 
Recently, Radicchi and Arenas \cite{radicchi2013abrupt} showed that the process of building independent network layers into a multiplex network---which is 
a specific type of multilayer interconnected network in which nodes replicate at each layer---undergoes a structural transition in the algebraic connectivity as interconnections are formed. 
Specifically, if $q$ is the interaction strength between the network layers, for $q<q_c$ these networks are structurally distinguishable 
(and the system behavior is not affected by their detailed topology but depends only on the interconnection structure), 
whereas, for $q>q_c$ the interconnected network functions as a whole (and topological effects do play a role). 
Later, Darabi Sahneh \ea \cite{darabi2015exact} found an exact solution for $q_c$. Moreover, they observed that the structural transition disappears 
when one of the network layers has vanishing algebraic connectivity: layers of such interconnected network topologies become indistinguishable, despite very weak coupling between them.
Mart\'in-Hern\'andez \ea \cite{martinez2014algebraic} further showed that, for a multiplex, there exists a critical number of diagonal interlinks beyond which 
any further inclusion does not enhance the algebraic connectivity of the system at all, whereas, for a randomly interconnected system, 
there exists a critical number of random interlinks beyond which algebraic connectivity increments at half of the original rate. 
Van Mieghem \cite{van2016interconnectivity} further computed the nontrivial eigenmode of the Laplacian for a regular topological structure of interconnections. 

Here we blend this research line of studying structural transitions in interacting networks. 
We adopt a perturbative approach in order to tackle general topologies of both network layers and interconnections. 
Perturbation theory has already found application in network science, for instance 
to study the Laplacian eigenvalues of scale-free networks \cite{kim2007ensemble}, 
to analyze spectral properties of networks with community structure \cite{chauhan2009spectral}, 
to identify important nodes within communities \cite{wang2011identifying}, 
to find the relation between eigenvector and topological perturbations \cite{yan2014eigenvector}, 
to analyze the localization properties of Laplacian eigenvectors on random networks \cite{hata2017localization} and, 
in the context of multiplex networks, to unveil the time scales of diffusive processes \cite{gomez2013diffusion,sole2013spectral}. 
The underlying idea of perturbation theory is to treat an operator acting on the system as the sum of an \emph{unperturbed} part, 
which in our context refers to isolated network layers and for which the exact solution may exist, 
and a \emph{perturbation}, given by the interconnections between these layers.

Our proposal constitutes a general framework for the analysis of structural transitions in the most wide scope of interconnected/interdependent multilayer networks. The analytical 
characterization of such transitions represents a step forward in the direction of having a closed theory of multilayer networks.

\section{Perturbative approach for the spectrum of the graph Laplacian}

We focus on studying the variation of the Laplacian matrix spectrum when the perturbation is introduced. 
We start with the simplest case of two connected, undirected unweighted networks $\mathcal{A}$ and $\mathcal{B}$, with $N$ and $M$ nodes each, respectively. 
Interconnections are randomly established between these networks, and are described by a generic $N\times M$ adjacency matrix $\mathcal{Q}$. 
The supra-Laplacian of the whole system can be represented with the four-blocks $(N+M)\times (N+M)$ matrix \cite{radicchi2013abrupt}:
\begin{equation}\label{eqn:L_firstdef}
\mathcal{L} = \left(
\begin{array}{cc}
\mathcal{L}_{\mathcal{A}} + \mathcal{K}_{\mathcal{A}} & -\mathcal{Q}\\ 
-\mathcal{Q}^{\mathsf{T}} & \mathcal{L}_{\mathcal{B}} + \mathcal{K}_{\mathcal{B}}
\end{array} \right),
\end{equation}
where $\mathcal{L}_{\mathcal{A}}$ and $\mathcal{L}_{\mathcal{B}}$ are the Laplacian matrices of each network, 
while $\mathcal{K}_{\mathcal{A}}=\mbox{diag}(\mathcal{Q}\ket{1})$ and $\mathcal{K}_{\mathcal{B}}=\mbox{diag}(\mathcal{Q}^{\mathsf{T}}\ket{1})$ are the diagonal matrices of inter-degrees. 
To apply perturbation theory, we split $\mathcal{L}$ into an unperturbed part $\mathcal{L}_0$ and a perturbation $\mathcal{V}$:
\begin{equation}\label{eqn:Lpert}
\mathcal{L} = \mathcal{L}_0 + \mathcal{V} =
\left(
\begin{array}{cc}
\mathcal{L}_{\mathcal{A}} & 0 \\
0 & \mathcal{L}_{\mathcal{B}}
\end{array} \right)+
\left(
\begin{array}{cc}
\mathcal{K}_{\mathcal{A}} & -\mathcal{Q} \\
-\mathcal{Q}^{\mathsf{T}} & \mathcal{K}_{\mathcal{B}}
\end{array} \right).
\end{equation} 
We denote, for $\mathcal{L}_0$, the unperturbed spectrum of eigenvalues as $E_n^{(0)}$ and its associated orthonormal basis of eigenvectors as $\ket{n^{(0)}}$. 
In the hypothesis of $E_n^{(0)}$ being non-degenerate, the first-order correction $\epsilon^{(1)}_n$ induced by the perturbation is
\begin{equation}\label{eqn:first_order}
\epsilon^{(1)}_n = \braket{n^{(0)}|\mathcal{V}|n^{(0)}},
\end{equation}
so that the spectrum of $\mathcal{L}$ at first order would be simply given by:
\begin{equation}\label{eqn:first_order_E1}
E^{(1)}_n = E^{(0)}_n + \braket{n^{(0)}|\mathcal{V}|n^{(0)}}.
\end{equation}

However we have to resolve the (at least) 2-fold degeneracy in the $0$ eigenvalue for $\mathcal{L}_0$, 
since there are at least two independent connected layers (networks $\mathcal{A}$ and $\mathcal{B}$). 
Assuming for simplicity that both $\mathcal{A}$ and $\mathcal{B}$ are connected, the degeneracy is exactly 2. 
We can then use the unperturbed eigenstates: 
\begin{eqnarray*}
\ket{+^{(0)}}&=&\frac{1}{\sqrt{N+M}}\binom{\ket{1}}{\ket{1}}\\
\ket{-^{(0)}}&=&\frac{1}{\sqrt{N+M}}\binom{\sqrt{\frac{M}{N}}\ket{1}}{-\sqrt{\frac{N}{M}}\ket{1}}
\end{eqnarray*}
as the orthonormal basis for such a degenerate sub-space \cite{van2016interconnectivity}. 
Since the perturbation $\mathcal{V}$ becomes diagonal when represented in this basis (\ie, $\braket{+^{(0)}|\mathcal{V}|-^{(0)}}=\braket{-^{(0)}|\mathcal{V}|+^{(0)}}=0$), 
we immediately get the eigenvalues corresponding to $\ket{+^{(0)}}$ and $\ket{-^{(0)}}$:
\begin{eqnarray}\label{eqn:1st_order_correction}
\epsilon^{(1)}_+ =&\braket{+^{(0)}|\mathcal{V}|+^{(0)}}&= 0, \\
\epsilon^{(1)}_- =&\braket{-^{(0)}|\mathcal{V}|-^{(0)}}&= \frac{\tau(\mathcal{Q})}{\mu},
\end{eqnarray}
where $\tau(\mathcal{Q}) = \braket{1|\mathcal{Q}|1} \equiv\sum_{ij}Q_{ij}$ and $\mu=NM/(N+M)$. 
Naturally, eq. \eqref{eqn:1st_order_correction} reminds of the classical two-body problem of two masses $N$ and $M$ mutually interacting 
by means of a coupling force of intensity $\tau(\mathcal{Q})$ \cite{van2016interconnectivity}: 
$\epsilon^{(1)}_+$ gives the acceleration for the center of mass while  $\epsilon^{(1)}_-$ is the relative acceleration between the two masses. 

We then consider the smallest non-zero eigenvalues of the unperturbed state $\mathcal{L}_0$ given by the algebraic connectivities of either network $\mathcal{A}$ or $\mathcal{B}$. 
Denoting as $\ket{v_{\mathcal{A}}}$ the normalized eigenvector corresponding to $\lambda_2(\mathcal{L}_{\mathcal{A}})$, 
that is $\mathcal{L}_{\mathcal{A}}\ket{v_{\mathcal{A}}}=\lambda_2(\mathcal{L}_{\mathcal{A}})\ket{v_{\mathcal{A}}}$, 
we pose $\ket{v_{\mathcal{A}}^{(0)}}=\binom{\ket{v_{\mathcal{A}}}}{\ket{0}}$. The first order correction to $\lambda_2(\mathcal{L}_{\mathcal{A}})$ is, according to eq. \eqref{eqn:first_order}: 
\begin{equation}
\epsilon^{(1)}_{\mathcal{A}} = \braket{v_{\mathcal{A}}^{(0)}|\mathcal{V}|v_{\mathcal{A}}^{(0)}}\equiv\braket{v_{\mathcal{A}}|\mathcal{K}_{\mathcal{A}}|v_{\mathcal{A}}}.
\end{equation}
Analogously, denoting as $\ket{v_{\mathcal{B}}}$ the normalized eigenvector corresponding to $\lambda_2(\mathcal{L}_{\mathcal{B}})$, 
and posing $\ket{v_{\mathcal{B}}^{(0)}}=\binom{\ket{0}}{\ket{v_{\mathcal{B}}}}$, we have: 
\begin{equation}
\epsilon^{(1)}_{\mathcal{B}} = \braket{v_{\mathcal{B}}^{(0)}|\mathcal{V}|v_{\mathcal{B}}^{(0)}}\equiv\braket{v_{\mathcal{B}}|\mathcal{K}_{\mathcal{B}}|v_{\mathcal{B}}}.
\end{equation}
Hence, the first order correction to the algebraic connectivity of $\mathcal{A}$ and $\mathcal{B}$ is given only by the degree configuration of the perturbation term
projected on the Fiedler vector of $\mathcal{L}_{\mathcal{A}}$ and $\mathcal{L}_{\mathcal{B}}$ respectively, independently on the particular topology of this perturbation term. 

Overall, at first order in perturbation theory we have: 
\begin{widetext}
\begin{equation}\label{eqn:lambda_firstorder_tot}
\lambda_2(\mathcal{L}) = \min \left\{ \frac{\tau(\mathcal{Q})}{\mu}, \;
\lambda_2(\mathcal{L}_{\mathcal{A}})+\braket{v_{\mathcal{A}}|\mathcal{K}_{\mathcal{A}}|v_{\mathcal{A}}}, \;
\lambda_2(\mathcal{L}_{\mathcal{B}})+\braket{v_{\mathcal{B}}|\mathcal{K}_{\mathcal{B}}|v_{\mathcal{B}}} \right\}.
\end{equation}
\end{widetext}
Since $\epsilon^{(1)}_-$ is the correction to the zero eigenvalue, we have that if $\tau(\mathcal{Q})$ is small enough then $\lambda_2(\mathcal{L}) = \epsilon^{(1)}_-$. 
In this phase the algebraic connectivity depends only on the sizes of the two interacting networks $\mathcal{A}$ and $\mathcal{B}$, meaning that it is {\em not affected} by their topology. 
However, when $\tau(\mathcal{Q})$ grows, the second and third smallest eigenvalues of the interacting network might swap \cite{martinez2014algebraic}. 
This happens when $\tau(\mathcal{Q})/\mu=\min \left\{\lambda_2(\mathcal{L}_{\mathcal{A}})+\braket{v_{\mathcal{A}}|\mathcal{K}_{\mathcal{A}}|v_{\mathcal{A}}},
\lambda_2(\mathcal{L}_{\mathcal{B}})+\braket{v_{\mathcal{B}}|\mathcal{K}_{\mathcal{B}}|v_{\mathcal{B}}} \right\}$.
Note that if one of the networks $\mathcal{A}$ and $\mathcal{B}$ has a vanishing algebraic connectivity, the transition point disappears \cite{van2016interconnectivity}. 
This happens, \eg, for a class of \SF networks where $\lambda_2(\mathcal{L}_{\mathcal{A}})\sim (\ln N)^{-2}$ \cite{samukin2008laplacian}. 
Importantly, an additional swapping may also occur for the algebraic connectivities of the two network layers, \ie, when and if 
$\lambda_2(\mathcal{L}_{\mathcal{A}})+\braket{v_{\mathcal{A}}|\mathcal{K}_{\mathcal{A}}|v_{\mathcal{A}}}=\lambda_2(\mathcal{L}_{\mathcal{B}})+\braket{v_{\mathcal{B}}|\mathcal{K}_{\mathcal{B}}|v_{\mathcal{B}}}$. 
To get a qualitative insight on the system behavior, in the following we consider two particular situations, diagonal and random interactions.

\subsection{Diagonal interactions (Multiplex)}

In a multiplex networks, $\mathcal{A}$ and $\mathcal{B}$ have the same number of nodes ($N=M$) and $\mathcal{Q}=q\mathcal{I}$ is proportional to the $N\times N$ identity matrix. 
While the minimization problem of eq. \eqref{eqn:lambda_courant} can be solved exactly in this case \cite{radicchi2013abrupt,darabi2015exact}, using perturbation theory leads to:
\begin{equation}\label{eqn:lambda_firstorder_multiplex}
\lambda_2(\mathcal{L}) = q+\min \left\{ q, \; \lambda_2(\mathcal{L}_{\mathcal{A}}),\; \lambda_2(\mathcal{L}_{\mathcal{B}}) \right\}.
\end{equation}
Since the ordering of $\lambda_2(\mathcal{L}_{\mathcal{A}})$ and $\lambda_2(\mathcal{L}_{\mathcal{B}})$ is fixed, 
there is only one eigenvalue swapping at $q_c\simeq \min \left\{ \lambda_2(\mathcal{L}_{\mathcal{A}}),\; \lambda_2(\mathcal{L}_{\mathcal{B}}) \right\}$. 

When the two networks $\mathcal{A}$ and $\mathcal{B}$ are identical, then $\lambda_2(\mathcal{L}_{\mathcal{A}})=\lambda_2(\mathcal{L}_{\mathcal{B}})$. 
Resolving this additional degeneracy with eigenvectors $\frac{1}{\sqrt{2}}\binom{\ket{v_{\mathcal{A}}}}{\ket{v_{\mathcal{A}}}}$ and $\frac{1}{\sqrt{2}}\binom{\ket{v_{\mathcal{A}}}}{-\ket{v_{\mathcal{A}}}}$ 
leads to first order corrections for $\lambda_2(\mathcal{L}_{\mathcal{A}})$ equal to $0$ and $2q$, hence $q_c\simeq \lambda_2(\mathcal{L}_{\mathcal{A}})/2$ 
(it is halved with respect to the non-degenerate case) \cite{martinez2014algebraic}. 

\subsection{Random interactions}

\begin{figure}
\centering
\includegraphics[width=8.6cm]{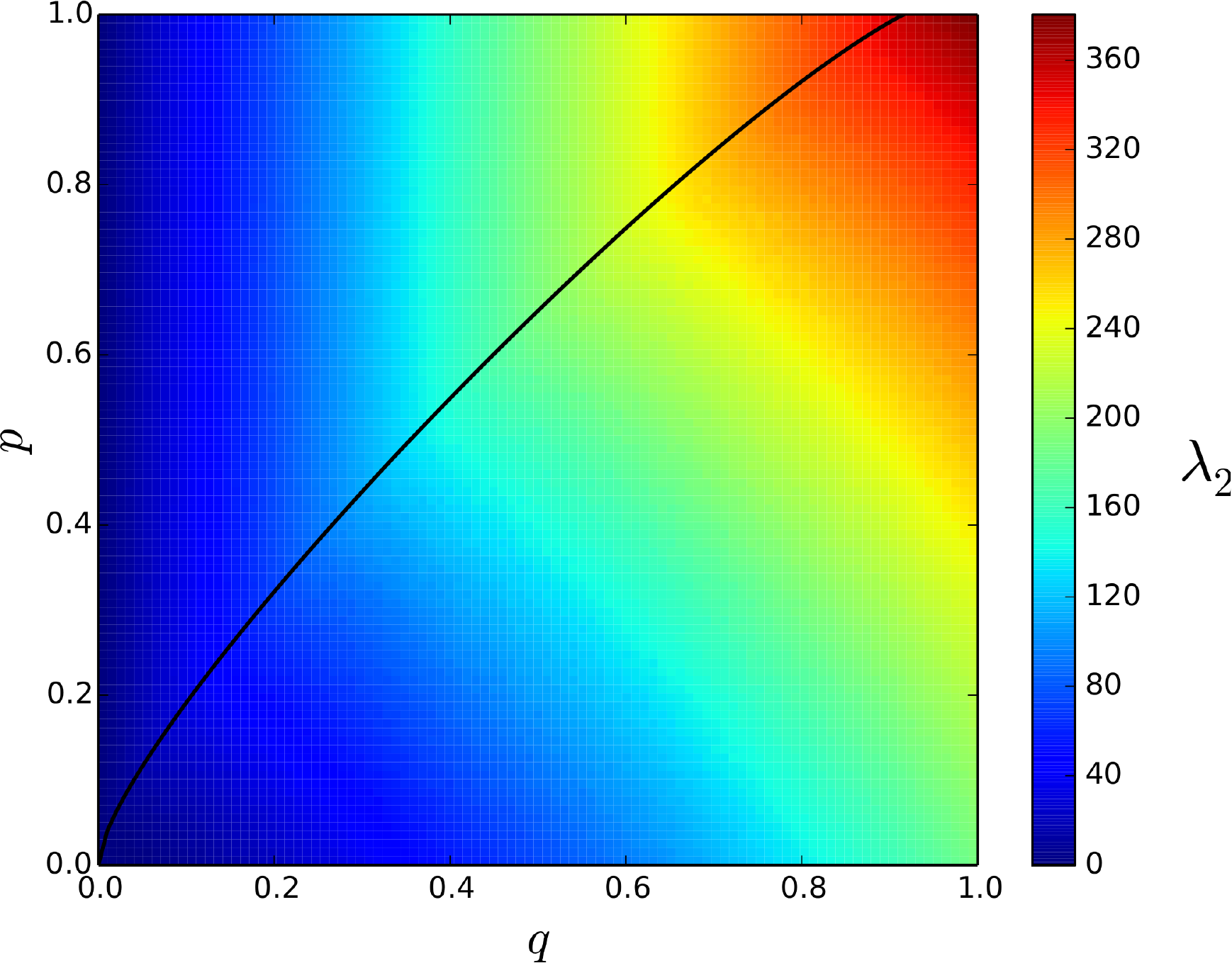}
\includegraphics[width=8.6cm]{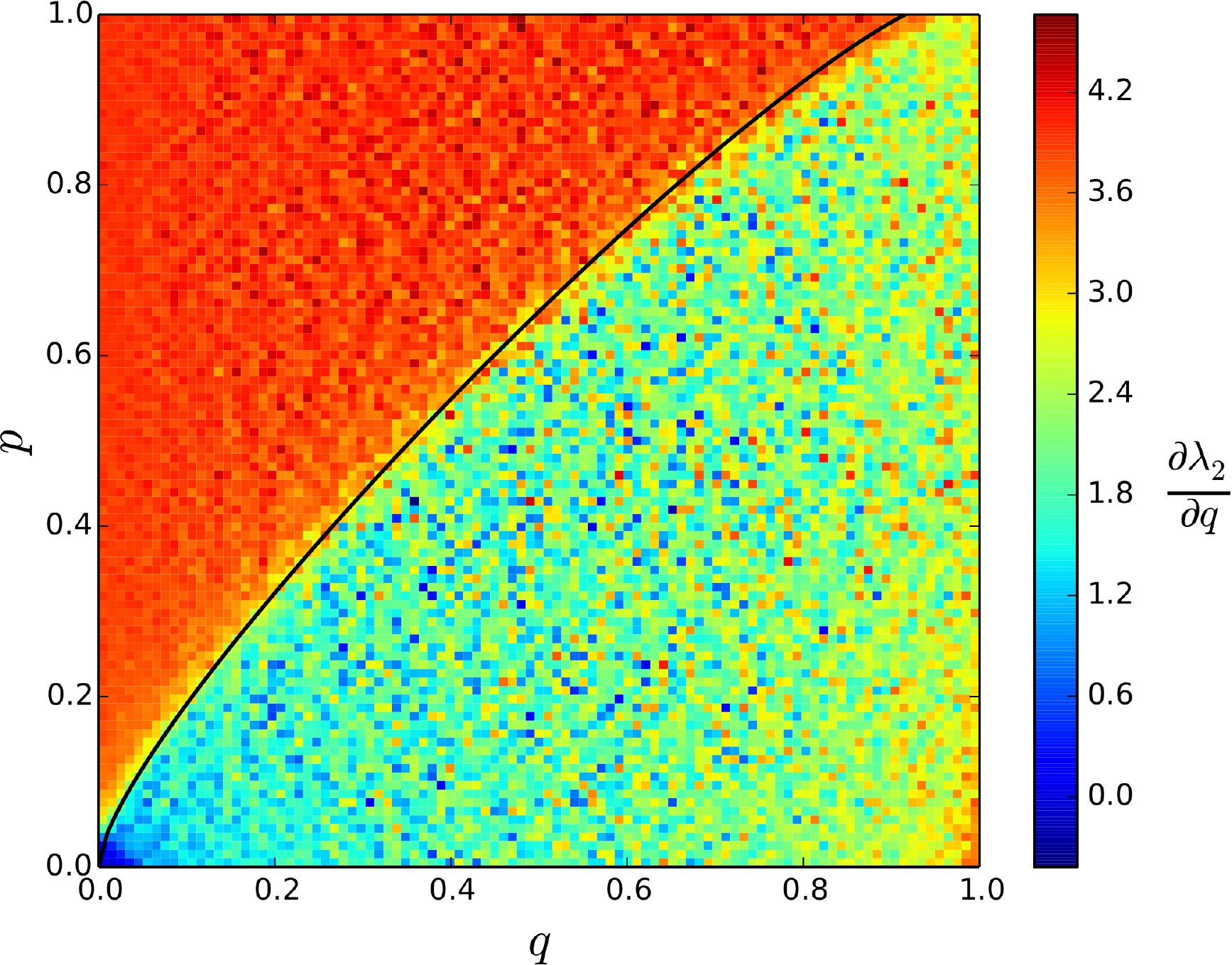}
\caption{Heat-map of $\lambda_2(\mathcal{L})$ (upper panel) and of $\partial\lambda_2(\mathcal{L})/\partial q$ (lower panel) for two interacting \ER graphs of $N=200$ nodes each 
and link probability $p$. The solid line is the curve described by eq. \eqref{eqn:border}.}\label{fig:heat}
\end{figure}

A more general situation is described by an interaction matrix $\mathcal{Q}$ assuming the form of an \ER random graph with connection probability $q$. 
This setting resembles that of an individual network with two communities $\mathcal{A}$ and $\mathcal{B}$ which are randomly interconnected \cite{chauhan2009spectral}. 
In order to proceed, we use a mean field approximation by replacing all matrix elements $Q_{ij}$ with their expectation value $q$. Hence eq. \eqref{eqn:lambda_firstorder_tot} becomes:
\begin{equation}\label{eqn:lambda_MF}
E[\lambda_2(\mathcal{L})] = \min \left\{ (N+M)q,  \lambda_2(\mathcal{L}_{\mathcal{A}})+qM,  \lambda_2(\mathcal{L}_{\mathcal{B}})+qN \right\}.
\end{equation}

\begin{figure*}[t]
\centering
\includegraphics[width=17.2cm]{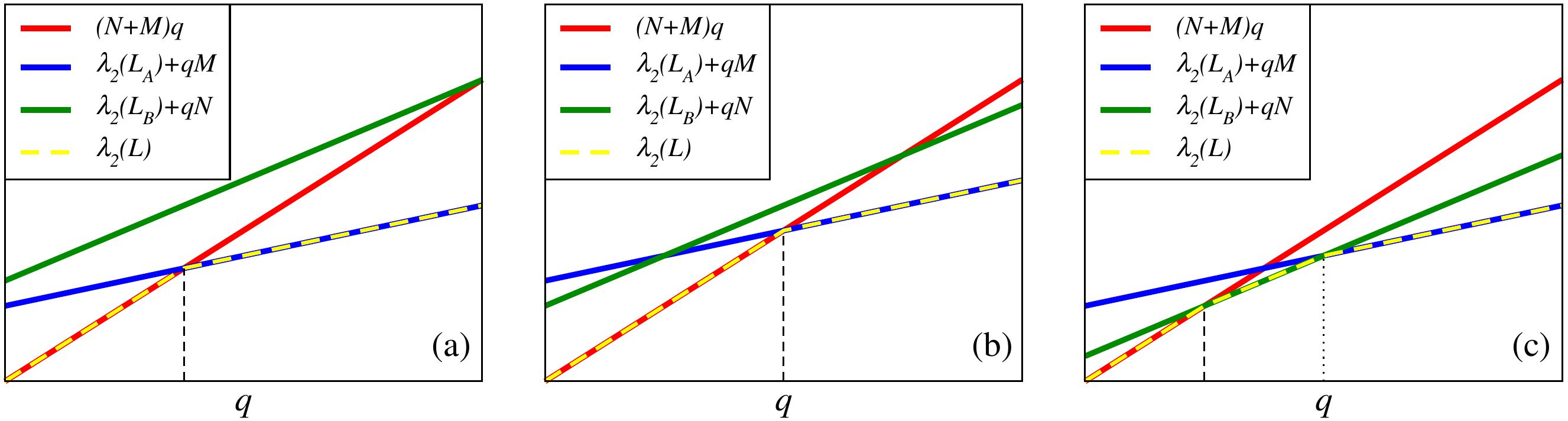}
\caption{Behavior of $\lambda_2(\mathcal{L})$ given by eq. \eqref{eqn:lambda_MF} for 
$M<N$ and: (a) $\lambda_2(\mathcal{L}_{\mathcal{A}})<\lambda_2(\mathcal{L}_{\mathcal{B}})$; 
(b) $\lambda_2(\mathcal{L}_{\mathcal{A}})>\lambda_2(\mathcal{L}_{\mathcal{B}})$ and $\lambda_2(\mathcal{L}_{\mathcal{A}})/N<\lambda_2(\mathcal{L}_{\mathcal{B}})/M$;
(c) $\lambda_2(\mathcal{L}_{\mathcal{A}})>\lambda_2(\mathcal{L}_{\mathcal{B}})$ and $\lambda_2(\mathcal{L}_{\mathcal{A}})/N>\lambda_2(\mathcal{L}_{\mathcal{B}})/M$.
%% actual params [ER, l=pN, N=100, M=N/2] (a): l_a=15N/100, l_b=4M/10; (b): l_a=N/5; l_b=3M/10; (c): l_a=15N/100, l_b=M/10.
Vertical dashed and dotted lines mark $q_c$ and $q_c'$, respectively.}\label{fig:cases}
\end{figure*}

Again in the special case of $\mathcal{A}$ and $\mathcal{B}$ identical (which also implies $N=M$), resolving the degeneracy $\lambda_2(\mathcal{L}_{\mathcal{A}})=\lambda_2(\mathcal{L}_{\mathcal{B}})$ 
with eigenvectors $\frac{1}{\sqrt{2}}\binom{\ket{v_{\mathcal{A}}}}{\ket{v_{\mathcal{A}}}}$ and $\frac{1}{\sqrt{2}}\binom{\ket{v_{\mathcal{A}}}}{-\ket{v_{\mathcal{A}}}}$ leads to first order corrections 
both equal to $Nq$, so that also in this case there is only one eigenvalue swapping at $q_c\simeq \lambda_2(\mathcal{L}_{\mathcal{A}})/N$ \cite{martinez2014algebraic}. 
Under the mean-field approximation, these conclusions hold also if the two networks are identical on expectation. 
For instance, consider $\mathcal{A}$ and $\mathcal{B}$ to be \ER random graphs with the same number of nodes and connection probability $p$. 
Dropping terms below $O(\sqrt{N\log{N}})$, we have $E[\lambda_2(\mathcal{L}_{\mathcal{A}})]=E[\lambda_2(\mathcal{L}_{\mathcal{B}})]\simeq Np-\sqrt{2p(1-p)N\log{N}}$ 
\cite{bollobas1981degree,jamakovic2008robustness}, hence
\begin{equation}\label{eqn:border}
q_c \approx p - \sqrt{2p(1-p)(\log{N})/N}
\end{equation}
(see Fig. \ref{fig:heat}). In the limit $N\to\infty$, $q_c\to p$ as $\sqrt{\log{N}/N}$: the transition at $q_c$ is therefore well defined even in the thermodynamic limit. 

\begin{figure}
\centering
\includegraphics[width=8.6cm]{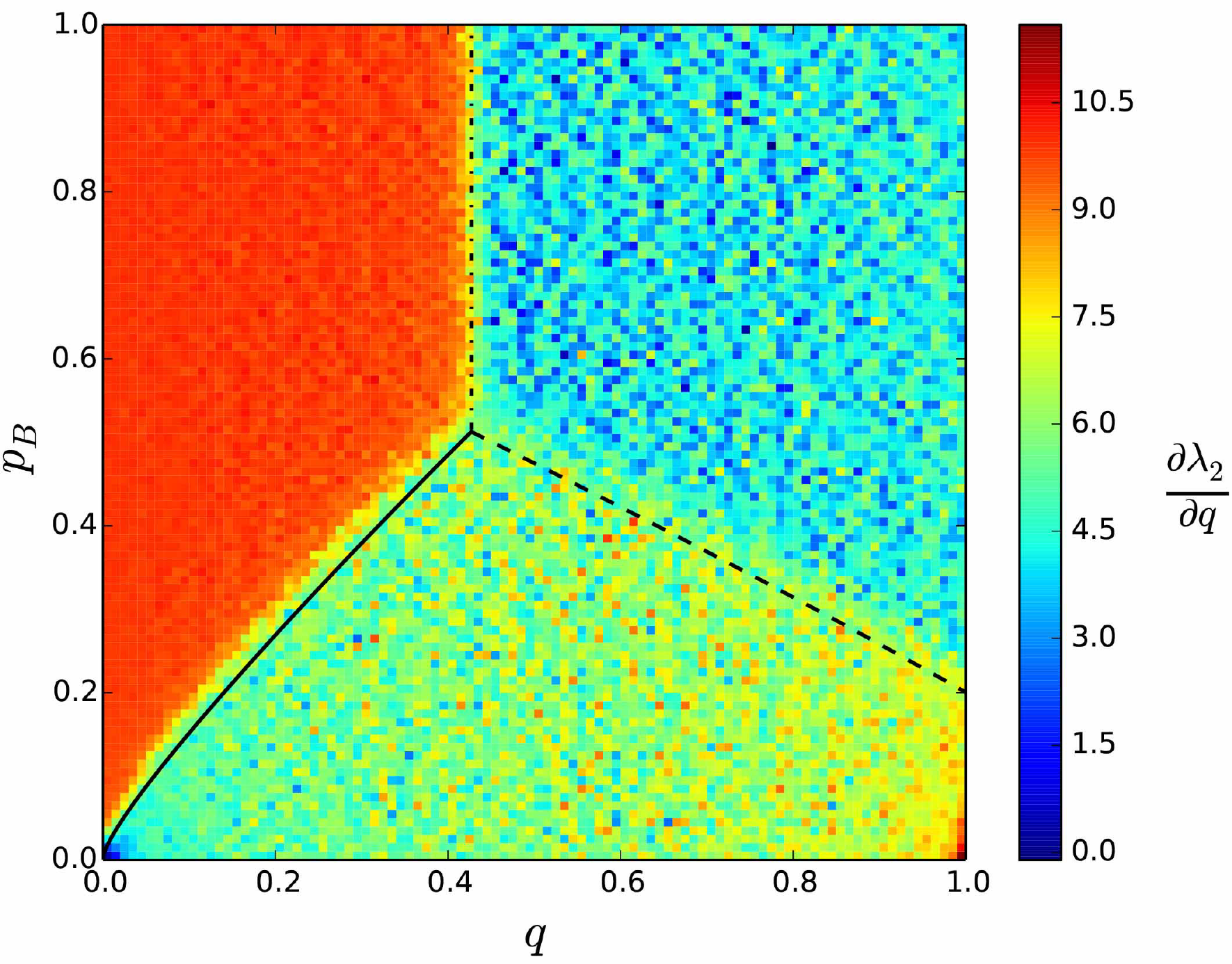}
\caption{Heat-map of $\partial\lambda_2(\mathcal{L})/\partial q$ for two interacting \ER graphs of $N=600$, $p_A=0.5$, $M=400$ and varying $p_B$. 
The three phases are delimited by the curve of eq. \eqref{eqn:border} with $p=p_B$ (solid line) and with $p=p_A$ (dashed-dotted line), plus the curve of eq. \eqref{eqn:q_II} (dashed line). 
Indeed, for $p_B\gtrsim 0.51(7)$ we are in the case $\lambda_2(\mathcal{L}_{\mathcal{A}})/N<\lambda_2(\mathcal{L}_{\mathcal{B}})/M$, 
hence there is only one transition at $q_c=\lambda_2(\mathcal{L}_{\mathcal{A}})/N\simeq 0.402(5)$. 
Instead for $p_B\lesssim 0.51(7)$ the first transition lies at $q_c=\lambda_2(\mathcal{L}_{\mathcal{B}})/M$, 
and the second one at $q_c'=[\lambda_2(\mathcal{L}_{\mathcal{A}})-\lambda_2(\mathcal{L}_{\mathcal{B}})]/(N-M)$ as long as $p_B\gtrsim 0.19(5)$. 
The triple point lies at $q_t=\lambda_2(\mathcal{L}_{\mathcal{A}})/N=\lambda_2(\mathcal{L}_{\mathcal{B}})/M$.}\label{fig:heat3}
\end{figure}

To discuss the more general setting of $\mathcal{A}$ and $\mathcal{B}$ having different sizes and topologies, 
without loss of generality we set $M<N$. Then if $\lambda_2(\mathcal{L}_{\mathcal{A}})<\lambda_2(\mathcal{L}_{\mathcal{B}})$, 
eq. \eqref{eqn:lambda_MF} tells us that the algebraic connectivity of $\mathcal{A}$ grows at a slower rate than that of $\mathcal{B}$, and they never become equal: 
only one eigenvalue swapping is possible, occurring again at
\begin{equation}\label{eqn:q_I}
q_c = \lambda_2(\mathcal{L}_{\mathcal{A}})/N.
\end{equation}
Instead if $\lambda_2(\mathcal{L}_{\mathcal{A}})>\lambda_2(\mathcal{L}_{\mathcal{B}})$, the first eigenvalue swapping occurs at 
$q_c =\min\left\{\lambda_2(\mathcal{L}_{\mathcal{A}})/N,\lambda_2(\mathcal{L}_{\mathcal{B}})/M\right\}$. 
Moreover, also the two algebraic connectivities of $\mathcal{A}$ and $\mathcal{B}$ swap at
\begin{equation}\label{eqn:q_II}
q_c' = \frac{\lambda_2(\mathcal{L}_{\mathcal{A}})-\lambda_2(\mathcal{L}_{\mathcal{B}})}{N-M}. 
\end{equation}
Such a transition is actually observed for $\lambda_2(\mathcal{L})$ only when $q_c<q_c'$, implying $\lambda_2(\mathcal{L}_{\mathcal{A}})/N>\lambda_2(\mathcal{L}_{\mathcal{B}})/M$ 
and when $q_c'<1$, implying $\lambda_2(\mathcal{L}_{\mathcal{A}})+M<\lambda_2(\mathcal{L}_{\mathcal{B}})+N$. Figure \ref{fig:cases} illustrates the different situations. 
Note that the second transition happens even for $\lambda_2(\mathcal{L}_{\mathcal{B}})\to0$ (\ie, when $\mathcal{B}$ is a \SF network): $q_c\to0$ but $q_c'\neq0$, 
provided $\lambda_2(\mathcal{L}_{\mathcal{A}})$ remains finite yet smaller than $N-M$. 
The phase diagram of Fig. \ref{fig:heat3} refers instead to $\mathcal{A}$ and $\mathcal{B}$ being \ER random graphs with connection probabilities $p_A$ and $p_B$ respectively. 
In the thermodynamic limit and for $r=M/N<1$ finite, for $p_A<p_B$  one transitions is observed at $q_c\simeq p_A$, whereas, for $p_A>p_B$ two transitions are observed 
at $q_c\simeq p_B$ and $q_c'\simeq(p_A-rp_B)(1-r)$, provided $q_c'<1$. The triple point obtains at $p_A=p_B=q$, \ie, when the whole system is homogeneous.

\begin{figure}
\centering
\includegraphics[width=8.6cm]{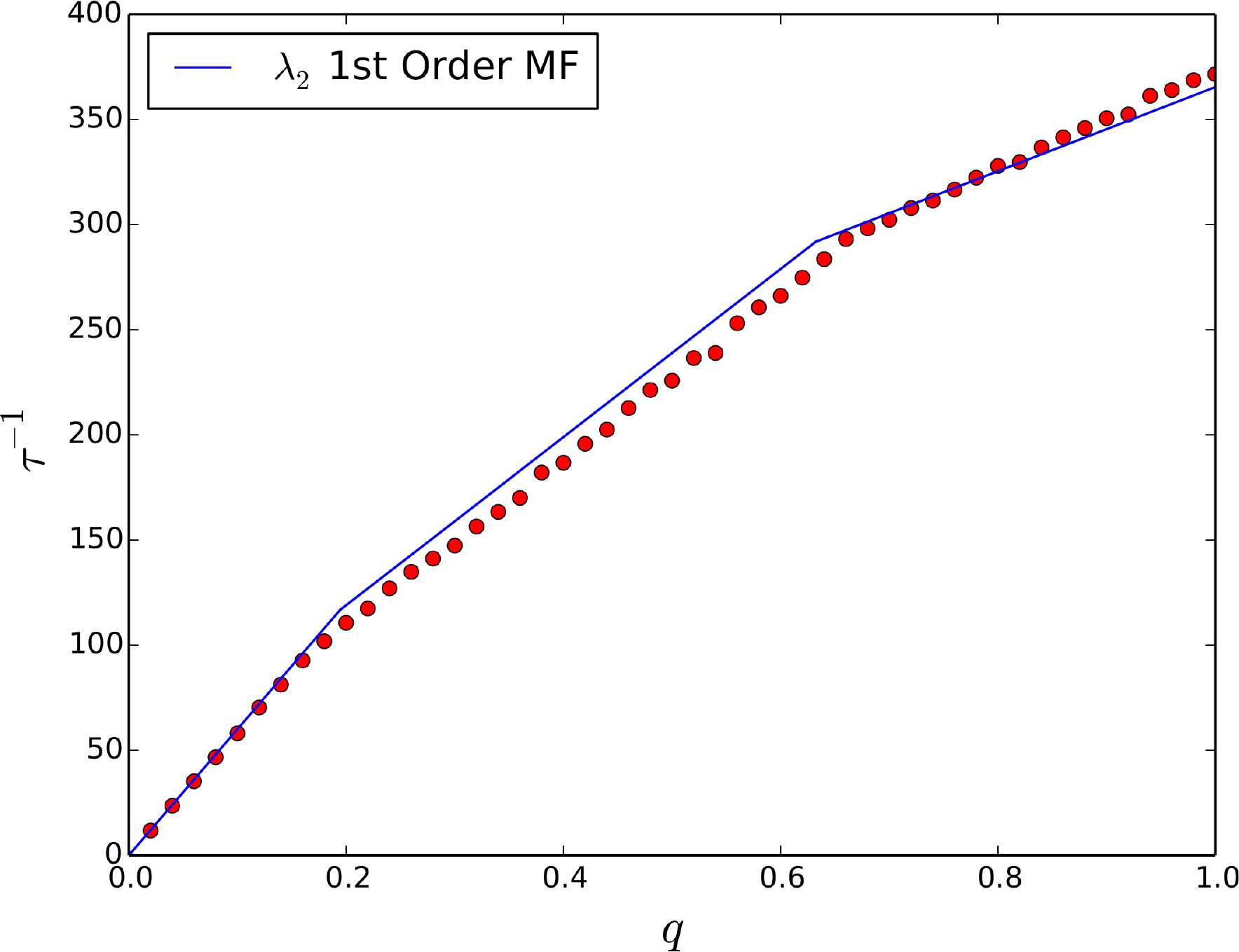}
\caption{Inverse relaxation time $\tau^{-1}$ for the diffusion process $\dot{\vec{\textbf{x}}} = -\mathcal{L}\vec{\textbf{x}}$ 
on two \ER randomly interconnected networks with $N=450$, $p_A = 0.45$, $M=300$, $p_B = 0.3$ and varying $q$. 
Red points refer to numerical simulations, whereas, the blue solid line indicates the first order mean field approximation of $\lambda_2(\mathcal{L})$ of eq. \eqref{eqn:lambda_MF}.}\label{fig:tau}
\end{figure}

The double transition of the algebraic connectivity described above is extremely important in the context of diffusion processes, 
since $\lambda_2^{-1}(\mathcal{L})$ is equal to the \emph{relaxation time} $\tau$ for the diffusion equation $\dot{\vec{\textbf{x}}}=-\mathcal{L}\vec{\textbf{x}}$ \cite{delvenne2015diffusion,masuda2017random}.
In the regime of small $q$, diffusion on the system depends only on the interconnection structure. 
The first transition occurs when the layer with the smallest normalized algebraic connectivity (be it $\lambda_2(A)/N$ or $\lambda_2(B)/M$) starts determining the diffusion process. 
The second transition then occurs when the other layer becomes dominant, and can be observed because the two algebraic connectivities grow at different rates ($N\neq M$) as $q$ increases. 
Note that the system becomes completely homogeneous only at the triple point $q_t$, when neither $A$ nor $B$ nor interconnections are dominant. 
Figure \ref{fig:tau} shows that values of $\tau^{-1}$ obtained from numerical simulations of such diffusion processes 
on random interacting networks do agree well with first order mean field approximation of $\lambda_2$.

\section{Perturbative approach for the spectrum of the adjacency matrix}

We now apply perturbation theory in the computation of the leading eigenvalue of the adjacency matrix of two interacting networks, which can be decomposed as:
\begin{equation}\label{eqn:Apert}
\mathcal{C} = \mathcal{C}_0 + \mathcal{W} =
\left(
\begin{array}{cc}
\mathcal{A} & 0 \\
0 & \mathcal{B}
\end{array} \right)+
\left(
\begin{array}{cc}
0 & \mathcal{Q} \\
\mathcal{Q}^{\mathsf{T}} & 0
\end{array} \right).
\end{equation} 
We denote by $\{\alpha_i\}_{i=1}^N$ and $\{\ket{a_i}\}_{i=1}^N$ the set of eigenvalues and eigenvectors of $\mathcal{A}$, ordered such that $\alpha_1>\alpha_2\ge\dots\ge\alpha_N$, 
and by $\{\beta_j\}_{j=1}^M$ and $\{\ket{b_j}\}_{j=1}^M$ the set of eigenvalues and eigenvectors of $\mathcal{B}$, again ordered such that $\beta_1>\beta_2\ge\dots\ge\beta_M$.
We assume both networks to be strongly connected, so that both $\alpha_1$ and $\beta_1$ are not degenerate in their respective spectrum. 
We also suppose, without loss of generality, $\alpha_1\ge\beta_1$. 
The sets $\{\Gamma_c^{(0)}\}_{c=1}^{N+M}=\{\{\alpha_i\}_{i=1}^N,\{\beta_j\}_{j=1}^M\}$ 
and $\{\ket{c^{(0)}}\}_{c=1}^{N+M}=\{\{\binom{\ket{a_i}}{\ket{0}}\}_{i=1}^N,\{\binom{\ket{0}}{\ket{b_j}}\}_{j=1}^M\}$ 
are thus the unperturbed spectrum of eigenvalues and its associated orthonormal basis of eigenvectors for $\mathcal{C}_0$. 

If $\alpha_1=\beta_1$, we have to resolve the degeneracy with the unperturbed eigenstates 
$\ket{+^{(0)}}=\frac{1}{\sqrt{2}}\binom{\ket{a_1}}{\ket{b_1}}$ and $\ket{-^{(0)}}=\frac{1}{\sqrt{2}}\binom{\ket{a_1}}{-\ket{b_1}}$. 
We have $\braket{+^{(0)}|\mathcal{W}|-^{(0)}}=\braket{-^{(0)}|\mathcal{W}|+^{(0)}}=0$, and:
\begin{eqnarray}\label{eqn:first_order_A_deg}
\gamma^{(1)}_+ =&\braket{+^{(0)}|\mathcal{W}|+^{(0)}}&= \braket{a_1|\mathcal{Q}|b_1}, \\
\gamma^{(1)}_- =&\braket{-^{(0)}|\mathcal{W}|-^{(0)}}&= -\braket{a_1|\mathcal{Q}|b_1}.
\end{eqnarray}
There is no degeneracy instead when $\alpha_1>\beta_1$. In this case, however, first-order corrections to all eigenvalues induced by the perturbation vanish: 
\begin{equation}\label{eqn:first_order_A}
\gamma^{(1)}_{c} = \braket{c^{(0)}|\mathcal{W}|c^{(0)}}=0,
\end{equation}
and we have to resort to second-order corrections. For $\alpha_1$ we have:
\begin{equation}\label{eqn:second_order_A}
\gamma^{(2)}_1 = \sum_{j=1}^M\frac{|\braket{a_1|\mathcal{Q}|b_j}|^2}{\alpha_1-\beta_j},
\end{equation}
where we used $\braket{b_j|\mathcal{Q}^{\mathsf{T}}|a_i}=\braket{a_i|\mathcal{Q}|b_j}$ $\forall i,j$. 
If also $\beta_1$ is non degenerate, then 
\begin{equation}\label{eqn:second_order_B}
\gamma^{(2)}_{N+1} = \sum_{i=1}^N\frac{|\braket{a_i|\mathcal{Q}|b_1}|^2}{\beta_1-\alpha_i}.
\end{equation}
It turns out, however, that second-order corrections fail to capture the behavior of $\Gamma_1$ (see Figure \ref{fig:adj}). 
In order to obtain a non-vanishing first-order correction, we have to define the unperturbed system and the perturbation as 
\begin{equation}\label{eqn:Apert_split}
\mathcal{C} = \tilde{\mathcal{C}}_0 + \tilde{\mathcal{W}} =
\left(
\begin{array}{cc}
\mathcal{A} & 0 \\
0 & \mathcal{B} + \Delta\mathcal{I}
\end{array} \right)
+
\left(
\begin{array}{cc}
0 & \mathcal{Q} \\
\mathcal{Q}^{\mathsf{T}} & -\Delta\mathcal{I}
\end{array} \right),
\end{equation}
where $\Delta = \alpha_1 - \beta_1$: we shift the whole unperturbed spectrum of $\mathcal{B}$ by $\Delta$, so that $\alpha_1$ is now a degenerate eigenvalue for $\tilde{\mathcal{C}_0}$ 
with respect to the same eigenvectors $\ket{a_1}$ and $\ket{b_1}$. Resolving the degeneracy with the same unperturbed eigenstates $\ket{+^{(0)}}$ and $\ket{-^{(0)}}$ as above, 
we obtain $\braket{+^{(0)}|\tilde{\mathcal{W}}|-^{(0)}}=\braket{-^{(0)}|\tilde{\mathcal{W}}|+^{(0)}}=\Delta/2$, 
$\braket{+^{(0)}|\tilde{\mathcal{W}}|+^{(0)}}=\braket{a_1|\mathcal{Q}|b_1}-\Delta/2$ and $\braket{-^{(0)}|\tilde{\mathcal{W}}|-^{(0)}}=-\braket{a_1|\mathcal{Q}|b_1}-\Delta/2$, hence
\begin{equation}\label{eqn:first_order_A_tilde}
\tilde{\gamma}^{(1)}_{\pm} = -\frac{\Delta}{2} \pm \sqrt{\frac{\Delta^2}{4} + [\braket{a_1 | \mathcal{Q} | b_1}]^2},
\end{equation}
which correctly reduces to eq. \eqref{eqn:first_order_A_deg} if $\Delta = 0$, and to $\tilde{\gamma}^{(1)}_{+}=0$ and $\tilde{\gamma}^{(1)}_{-}=-\Delta$ if $\mathcal{Q}$ vanishes (which is trivially correct).

All of the above formulas can be further specified for simple instances of the interaction matrix. 
For a multiplex network, $N=M$ and $\mathcal{Q}=q\mathcal{I}$, hence:
\begin{equation}\label{eqn:braket_multi}
\braket{a_i|\mathcal{Q}|b_j}=q\braket{a_i|b_j}.
\end{equation}
Instead for two randomly interacting networks, $\mathcal{Q}$ is an \ER random graph with connectivity $q$. 
Using the mean field approximation $\mathcal{Q}=q\ket{1}\bra{1}$ leads to:
\begin{equation}\label{eqn:braket_interact}
\braket{a_i|\mathcal{Q}|b_j}=q\braket{a_i|1}\braket{1|b_j}.
\end{equation}

\subsection{Random regular and \ER network layers}

More can be said when both $\mathcal{A}$ and $\mathcal{B}$ are $d$-regular graphs. 
In this case, it is $\alpha_1=d_A$, $\ket{a_1}=\frac{1}{\sqrt{N}}\ket{1}$, $\beta_1=d_B$, $\ket{b_1}=\frac{1}{\sqrt{M}}\ket{1}$. 
Besides, for sufficiently large network sizes, most $d$-regular graphs have all their other eigenvalues bounded above by $2\sqrt{d-1}+\varepsilon$ (with $\varepsilon>0$) \cite{friedman2003proof}. 
Thus, provided $d_B\gg 2\sqrt{d_A-1}$, $\alpha_1$ and $\beta_1$ are by far the largest eigenvalues of the unperturbed system. 
Finally, eigenvectors corresponding to other eigenvalues are orthogonal to $\ket{1}$, hence $\braket{1|a_i}=\braket{1|b_j}=0$ for $\neq 1$ and $j\neq 1$.

Thus in a multiplex framework where $N=M$ it is $\ket{a_1}\equiv\ket{b_1}$. Using eq. \eqref{eqn:braket_multi} and the eigenvectors orthogonality relations, 
we have $\braket{a_1|\mathcal{Q}|b_j}=q\delta_{1j}$ and $\braket{a_i|\mathcal{Q}|b_1}=q\delta_{i1}$. 
In the degenerate case we get $\gamma^{(1)}_{\pm}=\pm q$, whereas, in the non-degenerate case it is
$\tilde{\gamma}^{(1)}_{\pm} = -\Delta/2 \pm \sqrt{\Delta^2/4 + q^2}$ and $\gamma^{(2)}_1 =q^2/(\alpha_1-\beta_1)=-\gamma^{(2)}_{N+1}$.

In the random interaction framework instead, using eq. \eqref{eqn:braket_interact} and again the eigenvectors orthogonality relations, 
we have $\braket{a_1|\mathcal{Q}|b_j}=q\sqrt{NM}\delta_{1j}$ and $\braket{a_i|\mathcal{Q}|b_1}=q\sqrt{NM}\delta_{i1}$. 
In the degenerate case we get $\gamma^{(1)}_{\pm}=\pm q\sqrt{NM}$, and in the non-degenerate one 
$\tilde{\gamma}^{(1)}_{\pm} = -\Delta/2 \pm \sqrt{\Delta^2/4 + q^2NM}$ and $\gamma^{(2)}_1 =q^2NM/(\alpha_1-\beta_1)=-\gamma^{(2)}_{N+1}$.

\begin{figure}
\centering
\includegraphics[width=8.6cm]{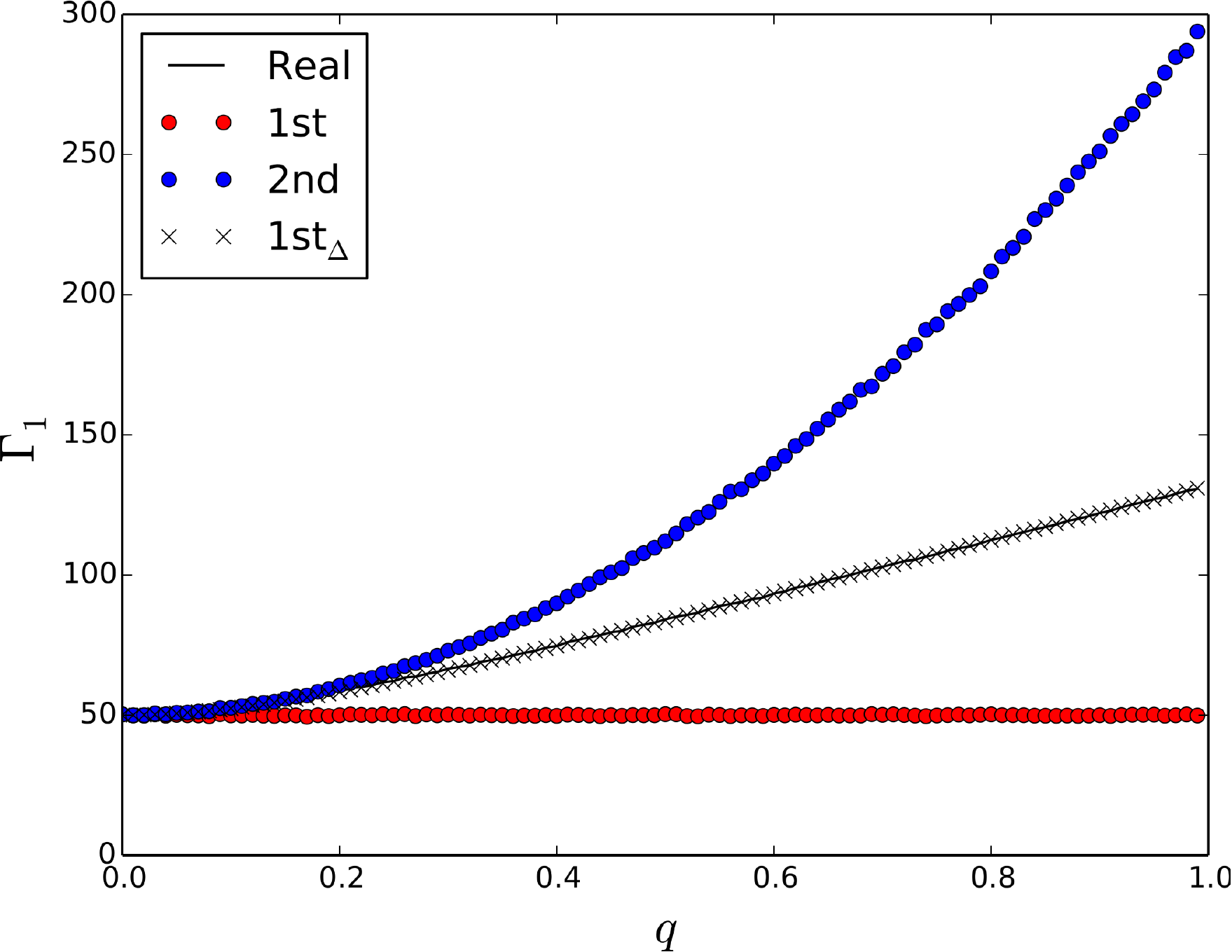}
\caption{Largest eigenvalue of $\mathcal{C}$ as a function of $q$ for two interacting \ER graphs of $N=M=100$, $p_A=0.5$ and $p_B=0.1$, 
together with first-order corrections of eq. \eqref{eqn:first_order_A}, second-order corrections of eq. \eqref{eqn:second_order_A}, 
and first-order corrections of eq. \eqref{eqn:first_order_A_tilde}.}\label{fig:adj}
\end{figure}

Finally note that a $d$-regular graph of size $N$ is, under the mean field approximation, equivalent to an \ER random graph with same size and connectivity $p=d/(N-1)$. 
Hence, the above results approximately hold also for $\mathcal{A}$ and $\mathcal{B}$ being \ER random graphs, in particular by posing $\alpha_1=p_A(N-1)$ and $\beta_1=p_B(M-1)$ (see Fig. \ref{fig:adj}).

\begin{figure}
\centering
\includegraphics[width=8.6cm]{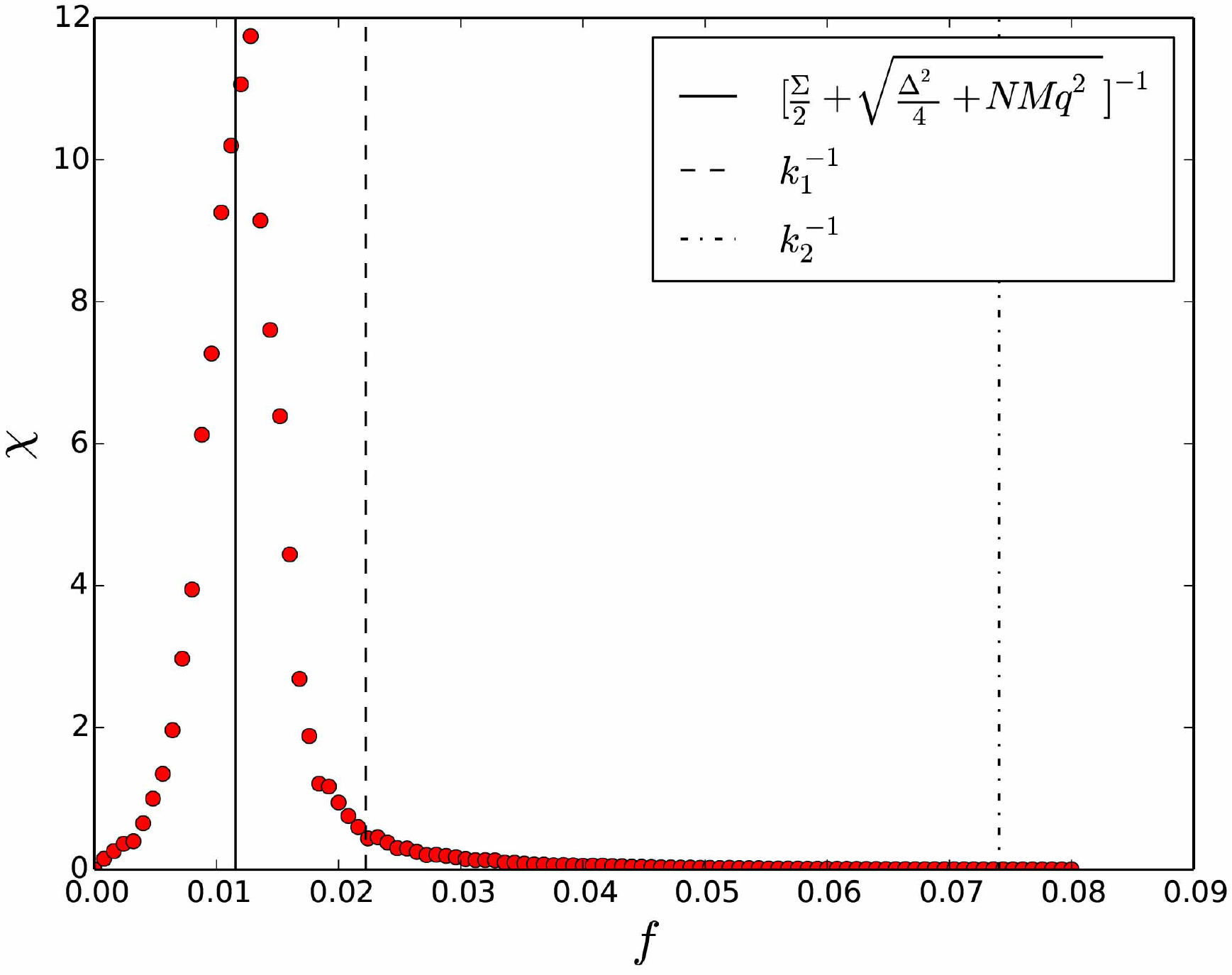}
\caption{Numerical value of the susceptibility $\chi$ as a function of the bond occupation probability $f$ for $400$ realizations of the process 
on two random ER interacting networks with $N=150$, $p_A = 0.1$, $M=100$, $p_B = 0.5$ and $q=0.5$. 
The solid black line denotes the mean field first-order correction to $\Gamma_1^{-1}$ of eq. \eqref{eqn:first_order_A_tilde}, 
whereas, the dashed and dashed-dotted lines denote the percolation thresholds of individual layers $k_1^{-1}=(Mp_B)^{-1}$ and $k_2^{-1}=(Np_A)^{-1}$. 
It is $\Sigma = k_1+k_2$ and $\Delta = k_1 - k_2$.}\label{fig:perc}
\end{figure}

This approach can be rather useful for estimating the bond percolation threshold $f_c$ of two strongly interacting random networks, where the magnitude of the interaction is given by the value of $q$. 
As a matter of fact when the value of $q$ is very small the two layers are in a regime of weak interaction, therefore two percolation thresholds are observed 
depending on the different topologies of the two layers \cite{hackett2016bond}\cite{simon2014doubleperc}.
On the other hand, while for an individual \ER network layer $f_c$ is given by the inverse of the largest eigenvalue of adjacency matrix $\Gamma_1$ \cite{bollobas2010percolation} 
(or in general is lower-bounded by $\Gamma_1^{-1}$ \cite{radicchi2016beyond}), Figure \ref{fig:perc} shows
that for two strongly interacting layers, where $q$ is not negligible, the percolation threshold is actually determined by eq. \eqref{eqn:first_order_A_tilde}.

\section*{Conclusions}

In this work we have presented a perturbative approach to study the connectivity properties for a general class of interacting multilayer networks. 
We generalized previous results \cite{radicchi2013abrupt,martinez2014algebraic,darabi2015exact} 
showing the presence of multiple structural transitions for interacting networks as interconnections are formed. 
This fact has a direct consequence on many physical dynamical systems which are governed by the laplacian spectrum, \eg, diffusive processes. 
We have shown that beyond the first eigenvalue crossing, there might be as much as $Z-1$ additional transitions, where $Z$ is the number of network layers. 
In each of these regimes, the relaxation time of a diffusive processes on the entire system is set by a single layer. 
We further show that, at first order in perturbation theory, the growth of the algebraic connectivity of each network layer depends only on the degree sequence of the interactions 
(projected on the respective Fiedler vector), and not on the actual interaction topology.
We finally show results of perturbation theory applied to the adjacency matrix of the interconnected network, 
which can be rather useful to identify percolation transitions on strongly interacting networks. 
Our findings have, therefore, important implications in the design of robust interconnected networked system, 
particularly when the functioning of the entire system crucially depends on one or a few network layers. 
Moreover, they allow to better understand diffusion of epidemics, habits adoption, information, opinions in our multilayer-structured societies. 
Overall, our results constitute a step forward to a better understanding of linear and nonlinear processes on top of interacting network structures, 
in the direction of having a closed mathematical theory of interacting multilayer networks.

\acknowledgments{A. A. acknowledges the Spanish MINECO, Grant No. FIS2015-71582-C2-1. A. A acknowledges funding also from ICREA Academia and the James S. McDonnell Foundation. 
G.C. and G.C. acknowledge support from the EU projects DOLFINS (640772), CoeGSS (676547), Shakermaker (687941) and SoBigData (654024).}

\end{document}